\documentclass[twocolumn,english,aps,showpacs]{revtex4}
\usepackage{ae,aecompl}
\usepackage{helvet}

\usepackage[T1]{fontenc}
\usepackage[latin9]{inputenc}
\usepackage{amsmath}
\usepackage{graphicx}
\usepackage{amssymb}

\makeatletter
\@ifundefined{textcolor}{}
{%
 \definecolor{BLACK}{gray}{0}
 \definecolor{WHITE}{gray}{1}
 \definecolor{RED}{rgb}{1,0,0}
 \definecolor{GREEN}{rgb}{0,1,0}
 \definecolor{BLUE}{rgb}{0,0,1}
 \definecolor{CYAN}{cmyk}{1,0,0,0}
 \definecolor{MAGENTA}{cmyk}{0,1,0,0}
 \definecolor{YELLOW}{cmyk}{0,0,1,0}
 }

\makeatother

\makeatother

\usepackage{babel}

\begin{document}

\title{Fractal spin structures as origin of 1/f magnetic noise in superconducting circuits}

\author{K. Kechedzhi$^{1}$, L. Faoro$^{2,1}$, and L. B. Ioffe$^{1}$ }

\affiliation{$^{1}$ Department of Physics and Astronomy,~Rutgers The State University
of New Jersey, 136 Frelinghuysen rd, Piscataway,~08854 New Jersey,
USA }

\affiliation{$^{\text{2}}$ Laboratoire de Physique Theorique et Hautes Energies,
CNRS UMR 7589, Universites Paris 6 et 7, 4 place Jussieu, 75252 Paris,
Cedex 05, France}

\pacs{85.25.Dq, 73.23.-b, 65.60.ah, 75.78.-n, 73.50.Td}
\begin{abstract}
We analyze recent data on the complex inductance of dc SQUIDs that
show 1/f inductance noise highly correlated with conventional 1/f
flux noise. We argue that these data imply a formation of long range
order in fractal spin structures. We show that these structures appear
naturally in a random system of spins with wide distribution of spin-spin
interactions. We perform numerical simulations on the simplest model
of this type and show that it exhibits $1/f^{1+\zeta}$ magnetization
noise with small exponent $\zeta$ and reproduces the correlated behavior
observed experimentally.
\end{abstract}
\maketitle
Despite recent progress, the origin of low-temperature ($T\lesssim1K$)
$1/f$ flux noise in superconducting circuits remains an open question
for nearly 30 years \cite{Koch1983}. A progress in this problem would
be important for a number of applications: noise suppression would
improve the sensitivity of SQUID magnetometers \cite{Koch1983,Wellstood1987}
and eliminate the dominant source of decoherence for flux \cite{Flux}
and phase \cite{Phase} qubits. The problem has two parts: the microscopic
origin of the degrees of freedom that produce the noise and the origin
of the interaction responsible for their dynamics.

It is now accepted that the noise is due to the electron spins
localized at the surfaces of the
superconductors~\cite{Phase,Clarke,FaoroIoffe}. The surface
localization is implied by the weak dependence of the noise
amplitude on the area of SQUIDs: variation of the loop area over 5
orders of magnitudes does not lead to a significant systematic
change in the noise amplitude \cite{Koch1983,Clarke}. The
approximate temperature independence of the noise in the range
$20mK\lesssim T\lesssim500mK$ suggests that the magnetic system
responsible for the noise must be characterized by very low energy
scales pointing to electron or nuclear spins. The nuclear spins can
be ruled out because $1/f$ flux noise persists up to 10MHz
\cite{qubits} which is much larger than the typical energy scale
associated with nuclear spins $f\lesssim1$kHz \cite{FaoroIoffe}.
Recent data \cite{Sendelbach2008,Moler2009} confirm that the flux
noise is due to electron spins situated close to the surface of
superconductors and show that these spins are characterized by the
surface density $n_{s}\sim5\times10^{17}$m$^{-2}$. Further support
for this conclusion is provided by density functional calculations
\cite{ClarkeDFT} which found that localized electronic levels are
likely to occur at strongly disordered metal-insulator (MI)
interfaces.

The dynamics of the electron spins that generate the noise remains
poorly understood. However, very recently, an experimental work by
S. Sendelbach et al.~\cite{Sendelback2009} discovered a highly
unusual feature of the magnetic noise in dc-SQUIDs at millikelvin
temperatures that severely limits possible mechanisms. It was found
that the noise in the SQUID inductance has also $1/f$ power spectrum
and that the inductance fluctuations are highly correlated with the
usual $1/f$ flux noise. Moreover, the correlation between the two
noises grows as the temperature is decreased and becomes of the
order of unity below $T\sim100mK$. This suggests a common underlying
mechanism producing the noise in both inductance and flux. Because
inductance and flux are even and odd under time inversion operation
respectively their cross-correlation implies broken time inversion
symmetry and the appearance of a long range magnetic order.

The formation of magnetic order is difficult to reconcile with the
temperature independence of the flux noise because normally the former
implies local fields of the order of the transition temperature, $T_{c}$,
that suppress individual spin dynamics at $T\ll T_{c}.$ The collective
modes are expected to be suppressed by anisotropy at the temperatures
in mK range so they cannot lead to temperature independent flux noise.
Similarly, the transitions between different metastable states in
ordered magnets involve domain wall motion which disappears at low
temperatures. Finally, the spin glass formation has to be ruled out
because one does not expect correlations between magnetization and
susceptibility in spin glasses, this expectation was confirmed by
simulations~\cite{Chen2010}.

\begin{figure}[b]
 \includegraphics[width=3in]{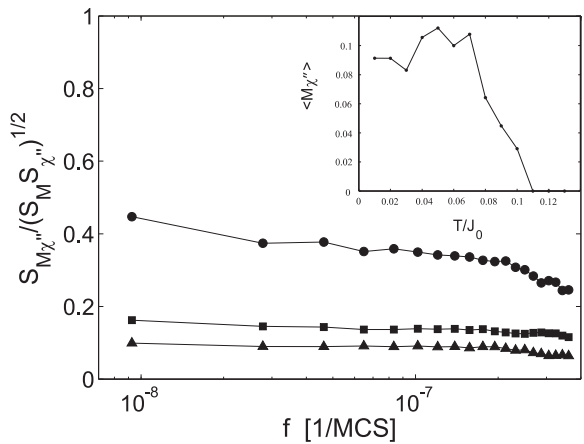}\caption{ Cross correlator of noise power spectra of the magnetization
and susceptibility for $T=0.05J_{0}$ and a particular disorder
configuration (circles); $T=0.05J_{0}$ averaged over disorder
(squares); $T=0.1J_{0}$ averaged over disorder (triangles). Inset
shows temperature dependence of the normalized disorder averaged
correlator of magnetization and susceptibility fluctuations $\langle
M\chi''\rangle$.}

\label{fig:1}
\end{figure}

In this paper we show that the experimental
results~\cite{Sendelback2009} are reproduced in a highly disordered
spin model where interaction between spins is characterized by a
very broad distribution of the coupling strengths. In this situation
the long range order is due to formation of a large fractal cluster
of strongly coupled spins which spans the whole system. This ordered
cluster includes only a fraction of spins leaving many smaller
clusters and isolated spins free to fluctuate. We formulate the
simplest model of this type and provide numerical results that agree
with the existing data. In more detail, we consider a system of
Ising spins distributed randomly in a 2D plane. Spin-spin
interactions are assumed ferromagnetic and decaying exponentially
with distance which provides a broad distribution of couplings
between them. We find that magnetization dynamics simulated by
single spin flip Monte Carlo algorithm produces noise with $1/f$
power spectra. We computed the linear response, $\chi(t)$, of the
system to a magnetic field of the low frequency, $\omega$, and
fluctuations of the magnetization of the system, $M(t)$, at time
scales much longer than the period of the external field,
$t\omega\gg1$. These quantities mimic the properties measured in the
experiments~\cite{Sendelback2009,IndSuscCorr}. Our main result is a
large cross-correlation between the noise in the susceptibility and
magnetization, as illustrated by Fig.~\ref{fig:1}, in a wide range
of temperatures in the ferromagnetic phase.

The model provides a new insight into the nature of spin-spin
interactions which is the most likely mechanism generating the
dynamics of the spins. First, we note that both phonons and nuclear
spins can be ruled out as a source of electron spin dynamics due to
their weak coupling to the electron spins~\cite{Clarke,FaoroIoffe}.
Alternative sources of the dynamics such as hopping of electrons
between traps with random spin orientations \cite{Clarke} or
interaction with tunneling two level systems (TLS) \cite{DeSouza}
are difficult to reconcile with experimental data. The former
requires very high concentration of single-electron excitations
while the latter implies high concentration of thermally activated
TLSs \cite{FaoroIoffe}. This leaves only electron spin interaction
with each other.

The typical spin-spin distance inferred from the reported densities
is $\gtrsim1nm$, which implies an exponentially small direct spin
exchange. However, proximity to the metal makes RKKY interaction
possible \cite{FaoroIoffe}. This interaction can be especially large
for the electrons localized on metal-insulator boundary levels that
remain singly occupied due to Coulomb repulsion \cite{ClarkeDFT},
while electrons localized deeper in the insulator are expected to
have exponentially weaker coupling to the conduction electrons. The
exponential dependence of the coupling to conduction electrons
implies a broad distribution of spin-spin couplings.

Theoretical analysis of a system of spins interacting via such random
RKKY mechanism under assumption of \emph{sufficiently weak} interactions
(such that the system does not form an ordered state) showed that
the system can generate temperature independent $1/f$ noise. Relatively
high frequency part of the noise spectra~\cite{Wellstood1987} is
generated by the diffusion of local magnetization~\cite{FaoroIoffe},
whereas the very low frequency part \cite{Phase,Lanting2009} is generated
by fluctuations of rare pairs of spins \cite{Faoro2011}.

Temperature independence of the $1/f$ noise and correlations between
flux and inductance~\cite{Sendelback2009} imply unusual magnetic
long range order. In a conventional magnet the noise is due to
domain wall jumps. Each such jump induces changes in the
magnetization and susceptibility proportional to the changes in the
domain volume $\delta V$ and domain wall area $\delta S$
respectively. This means that in conventional magnets where $S\ll V$
the fluctuations in the response to an external field is much
smaller then fluctuations in the quasistatic magnetization in
contrast to the data.

Large fluctuations of the response and as well as high correlations
between the response and magnetization are possible in highly
disordered magnets with wide distributions of spin-spin couplings
where correlated spins form fractal clusters with $V\sim S$. The
simplest model that contains this physics is characterized by the
Hamiltonian
\begin{equation} \mathcal{H}=-\sum_{ij}J_{ij}s_{i}s_{j},\;
J_{ij}=J_{0}\exp(-r_{ij}/a),\label{model}\end{equation}
 where $i$ numbers sites of the random lattice, $a$ is a decay length
of interactions, $J_{0}$ determines the annealing temperature of the
system and spins are classical $s=\pm1$. Three dimensional version
of this model was used earlier to analyze the formation of long
range order in magnetically doped
semiconductors~\cite{Shender,DasSarma}.

At small densities $na^{2}\ll1$ the system (\ref{model}) undergoes
finite temperature ferromagnetic transition which is driven either
by the temperature $T$ or by the lattice site density $n$. This
transition is in the universality class of percolation transition.
It can be understood in terms of the \textquotedbl{}circle
packing\textquotedbl{} problem in 2D in which one considers circles
of radius, $r(T)/2\equiv a/2\ln J_{0}/T$, drawn around each spin
site. Overlapping circles correspond to spins separated by
$r_{ij}\lesssim r(T)$ which are therefore strongly interacting and,
thus, aligned. In contrast, spins separated by $r_{ij}>r(T)$ are
effectively independent. Dimensionless parameter $B(T)=\pi
r^{2}(T)n$ controls the thermodynamic transition. At high
temperatures $B(T)\ll1$ strongly coupled spins are rare and form
only small clusters. As the temperature decreases, $B(T)$ increases,
both the number and sizes of the clusters grow and at some threshold
value $B=B_{c}$ an infinite cluster of strongly coupled spins forms.
This corresponds to the ferromagnetic ordering at the critical
temperature, $T_{c}=J_{0}\exp\left(-\sqrt{B_{c}/\left(\pi
na^{2}\right)}\right)$. The threshold value for the
\textquotedbl{}circle packing\textquotedbl{} problem is
$B_{c}\approx4.5$ \cite{EfShkl}. In the vicinity of the peroclating
transition the cluster size distribution is given by the power law:
$n(s)\sim s^{-\tau}\exp(-Cs^{1/2})$, $B>B_c$, where $C$ is a
constant and $\tau$ is a characteristic exponent~\cite{EfShkl}. The
scaling behavior is realized for $|B-B_{c}|\ll B_{c}$ which
translates into a wide range of temperatures:
$\ln(T_{c}/T)\ll1/\sqrt{na^{2}}$.

\begin{figure}[t]
 \includegraphics[width=3in]{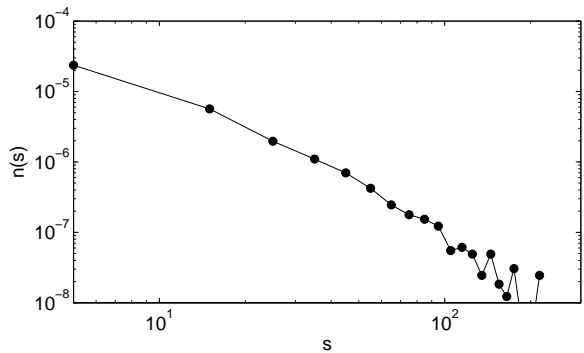}\caption{Cluster size distribution $n(s)$
 at $|B_{c}-B|/B_{c}=0.2$. Up to
sizes $s\gtrsim100$ the data fits a power law $n(s)\sim s^{-\tau}$
with $\tau\sim2$.}

\label{fig:2}
\end{figure}

We analyzed numerically the cluster size distribution in the model
(\ref{model}) in the parameter range $0\lesssim|B_{c}-B|/B_{c}\lesssim0.2$.
This translates into $|\ln T_{c}/T|\lesssim0.5$ for the model characterized
by realistic values of $\ln J_{0}/T_{c}\approx5$. We found a broad
power law distribution of cluster sizes up to $s\gtrsim100$ at $|B_{c}-B|/B_{c}=0.2$,
see Fig.~\ref{fig:2}, indicating that in this whole temperature
range the spin system separates into many large clusters weakly coupled
to each other and to the infinite percolation cluster.

Formation of large clusters of spins that flip as whole objects
under thermal excitation affect the magnetization dynamics of the
system for two reasons. First, the contribution to the noise coming
from large clusters scales as $n(s)s^{2}$ so their relatively small
number is compensated by the additional factor of $s$. Second, the
broad distribution of the cluster sizes translates into wide
distribution of relaxation rates $\Gamma$ associated with cluster
dynamics which results in $1/f$ power spectra of the magnetization
noise.

\begin{figure}[t]
 \includegraphics[width=3in]{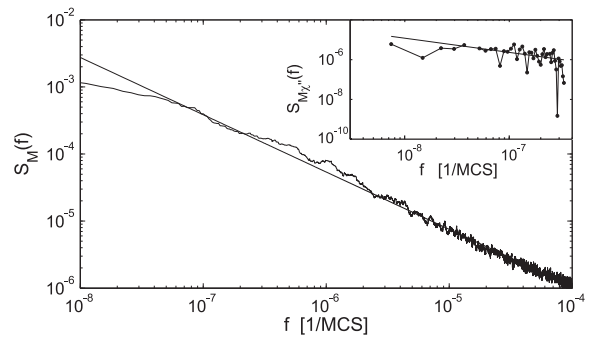} \caption{Normalized power spectra of the magnetization noise at $T=0.05J_{0}$
of the lattice with $N=1024$ sites. Linear fit yields $\alpha=0.93$
for the noise exponent. Inset shows $S_{M\chi''}(f)$ at $T=0.07J_{0}$,
$\alpha=0.72$.}

\label{fig:3}
\end{figure}

In order to check this conjecture we have simulated the dynamics of
the spin system (\ref{model}) in the critical regime using single
spin flip Monte Carlo dynamics satisfying the detailed balance condition.
We focus on the fluctuations of magnetization in the regime below
the freezing temperature, $T_{c}$, at which the infinite cluster
is formed. To minimize the transient regime we choose for the initial
configuration that has spins on the infinite cluster aligned in one
direction whereas all others are distributed randomly. Very long runs
up to $3\times10^{10}$ Monte Carlo steps (MCS) were performed to
analyze long time dynamics in the steady state characterized by finite
average magnetization. We observe $1/f^{\alpha}$ power spectra of
the noise in magnetization of the system, Fig~\ref{fig:3}, with
the noise exponent $0.8\lesssim\alpha\lesssim1.2$.

In order to simulate the susceptibility noise measurement \cite{Sendelback2009},
we subjected the spin system to a time-dependent magnetic field $H=H_{0}\cos2\pi\omega t$
and simulated the lock-in detection by applying digital low-pass filter~\cite{filter}
to the fluctuation of magnetization $M(t)$ and its Fourier component
$M(t)*\cos(2\pi\omega t+\phi)$, where $\phi=\pi/2$ corresponds to
the out-of-phase response $\chi''$. We take the field period to be
relatively long, $\omega^{-1}=3\times10^{5}$ MCS, for the system
of $N=1024$ spins. The resulting cross-correlator of the power spectra
$S_{M\chi''}(f)/(S_{M}(f)S_{\chi''}(f))^{1/2}$, where $S_{M}(f)\equiv\left(1/\overline{M^{2}}\right)\int_{0}^{\infty}dte^{-2\pi fti}\overline{M(0)M(t)}$,
is shown in Fig.~\ref{fig:1}. Temperature dependence of the cross-correlator
of the noise amplitudes $\langle M\chi''\rangle$ averaged over disorder
is shown in the inset of Fig.~\ref{fig:1}; power spectra of the
cross-correlator is shown in the inset of Fig.~\ref{fig:3}.

The noise spectrum and correlations observed in numerical simulations
can be understood using the following qualitative picture. Long time
scale dynamics responsible for the correlated noise in susceptibility
and magnetization is generated by flipping parts of infinite percolation
cluster and large clusters at its boundary. These flips are due to
thermal activation, so the rate of the cluster to flip is $\Gamma\sim\exp(-V/T)$,
where $V$ is the maximal energy barrier encountered on the optimal
spin flip path that flips the whole cluster.

\begin{figure}[t]
 \includegraphics[width=1.5in]{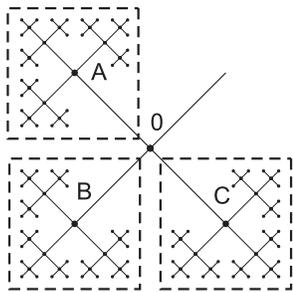}

\caption{Sketch of the fractal spin structure illustrating the logarithmic
dependence of the barrier with the cluster size. }

\label{Flo:Skecth}
\end{figure}

Fractal geometry of clusters results in a logarithmic dependence of
$V(s)$ on the cluster size. This unusual dependence was first seen
in numerical work~\cite{Rammal}; it can be understood theoretically
by assuming that clusters that dominate the dynamics at relatively
high frequencies are essentially random graphs~\cite{HenleyII}.
Locally, any such structure can be viewed as a tree growing from the
'origin' that is connected by $m+1$ bonds, of strength $J$, of which
$m$ connect to $m$ sites of the first level that are connected by
$m$ bonds each to another $m^{2}$ sites of the second level and so
on (see Fig.~\ref{Flo:Skecth} for $m=3$). The size of the cluster is
$1+(m^{L+1}-1)/(m-1)$ where $L$ is the number of levels. Denote by
$V_{A}=V_{B}=V_{C}=V$ the energy barrier to flip each of the
subtrees $A$, $B$ and $C$ in Fig.~\ref{Flo:Skecth}. An optimal path
to flip the whole tree involves flipping all subtrees in sequence
which means that the maximum barrier will be $V+J$. This barrier is
encountered when one of the subtrees, say $A$, have been flipped
which requires keeping a bond from \textquotedbl{}$0$\textquotedbl{}
to $A$ in the high energy state. Subsequent flipping of $B$ will
therefore require $V+J$. This is the maximum barrier encountered
because with the next step site \textquotedbl{}$0$\textquotedbl{}
can be flipped reducing the barrier for $C$ to $V+J$. In the case of
$m>3$ the maximum barrier will be $\approx V+m/2J$. This means that
the energy barrier grows linearly with the number of levels whereas
the size of the cluster is exponential. Therefore the resulting
dependence of the energy barrier on the size of the cluster is
logarithmic. One expects the logarithmic barrier scaling with size
to remain valid also in the case of a random tree characterized by
$m$ and $J$ fluctuating around average values $\langle m\rangle$ and
$\langle J\rangle$, which is a more accurate model. In the case of
percolation on the diluted Bethe lattice with fixed bond strength
$J=\langle J\rangle$
\begin{equation}
V(s)\approx\tfrac{Z\langle J\rangle}{D}\ln s,\label{blog}\end{equation}
 where $D$ is the fractal dimension of the tree and the parameter
$Z$ is restricted by $1/\ln2\leq Z\leq2/\ln2$~\cite{Henley}. A very
similar argument based on the 'links' and 'blobs' structure of the
cluster~\cite{Stanley} supports the validity of scaling (\ref{blog})
for the whole cluster~\cite{HenleyII}. We expect this logarithmic
dependence of the energy barriers on the cluster size should hold at
least approximately for the system of spins interacting via
exponential interaction (\ref{model}) at percolation threshold with
a similar value of $Z$.

The result (\ref{blog}) translates into the power law spectra of
the magnetization noise. At equilibrium the average autocorrelator
of the magnetic moment $M_{i}(t)$ of the $i$-th cluster is given
by $\overline{M_{i}(0)M_{i}(t)}\sim e^{-\Gamma_{i}t}$. Treating clusters
as independent and approximating the cluster distribution by $n(V,s)\approx\delta(V-V(s))s^{-\tau},$
we get \begin{gather}
S_{M}(f)\sim\int dsdVs^{2}n(s,V)\tfrac{\Gamma(V)}{\Gamma^{2}(V)+f^{2}}\sim\frac{1}{f^{1+\zeta(T)}},\label{1/feq}\\
\zeta(T)=\tfrac{T}{\langle J\rangle}\tfrac{D}{Z}(3-\tau),\nonumber \end{gather}
 At percolation threshold the values of the parameters $\tau=187/91$ and
$D=91/48$ are universal~\cite{Aharony}.

Out-of-phase response to the periodic magnetic field with relatively
high frequency $\omega$ is dominated by small clusters with $\Gamma\sim\omega$
that do not equilibrate during one field period. These clusters experience
a local quasistatic field created by all neighboring clusters that
fluctuate with frequencies $f\ll\omega$. When one of these clusters
flips the local field acting on the small cluster changes which results
in the change in its response to the external field at frequency $\omega.$
This leads to the susceptibility fluctuations which are correlated
with the magnetization fluctuations as observed in numerical simulations
and in the experiment.

In conclusion, we have shown that the main results of recent
experiments \cite{Sendelback2009} on the flux noise are reproduced
by the simplified model of a disordered ferromagnet with a very wide
distribution of couplings between spins. In the simulations we
assumed single spin flip dynamics which does not conserve local
spin. Understanding the mechanism producing such dynamics or
generalizing these results to other types of dynamics will be a
subject of future work.

We are grateful to Robert McDermott for patient explanations of his
data and useful discussions. This work was supported by ARO 56446-PH-QC,
ECS-0608842, DARPA HR0011-09-1-0009, Triangle de la Physique 2007-36.


\begin{thebibliography}{27}
\bibitem{filter} It is important to take into account in simulations
that low-pass filters used for lock-in detection in experiments preserve
causality of the signal. A digital filter acts on a discrete time
signal $m(t)$ as $\overline{m}(t)=\sum_{t'=0}h(t')m(t-t')$. The
causality condition leads to the following restrictions imposed on
the Fourier image $h(\omega)=\sum h(t)e^{-i\omega t}$~\cite{Proakis}:
(1) $\int_{-\pi}^{\pi}|\ln|h(\omega)||<\infty$; (2) real and imaginary
parts of $h(\omega)$ are related via $\Im h(\omega)=-\tfrac{1}{2\pi}
\int_{-\pi}^{\pi}d\nu\Re h(\nu)\cot\tfrac{\omega-\nu}{2}$.
\end{thebibliography}
\end{document}